\documentclass[amstex,aps,preprint,nofootinbib]{revtex4}
\usepackage[latin9]{inputenc}
\usepackage{textcomp}
\usepackage{amsmath}
\usepackage{amssymb}
\usepackage{graphicx}

\makeatletter
\@ifundefined{textcolor}{}
{%
 \definecolor{BLACK}{gray}{0}
 \definecolor{WHITE}{gray}{1}
 \definecolor{RED}{rgb}{1,0,0}
 \definecolor{GREEN}{rgb}{0,1,0}
 \definecolor{BLUE}{rgb}{0,0,1}
 \definecolor{CYAN}{cmyk}{1,0,0,0}
 \definecolor{MAGENTA}{cmyk}{0,1,0,0}
 \definecolor{YELLOW}{cmyk}{0,0,1,0}
}

\@ifundefined{textcolor}{}{%
 \definecolor{BLACK}{gray}{0}
 \definecolor{WHITE}{gray}{1}
 \definecolor{RED}{rgb}{1,0,0}
 \definecolor{GREEN}{rgb}{0,1,0}
 \definecolor{BLUE}{rgb}{0,0,1}
 \definecolor{CYAN}{cmyk}{1,0,0,0}
 \definecolor{MAGENTA}{cmyk}{0,1,0,0}
 \definecolor{YELLOW}{cmyk}{0,0,1,0}
}

\usepackage{color}
\usepackage{bm}
\usepackage{braket}
\usepackage{longtable}
\usepackage{amsfonts}
\usepackage{braket}

\makeatother

\begin{document}
\title{Nonlocal interferences induced by the phase of the 
	wavefunction
	for a particle in a cavity with moving boundaries}
\author{Mordecai Waegell}
\affiliation{Institute for Quantum Studies, Chapman University, Orange, CA 92866,
	USA}
	
\author{Alex Matzkin}
\affiliation{Laboratoire de Physique Th\'eorique et Mod\'elisation, CNRS Unit\'e 8089, 
    CY Cergy Paris Universit\'e, 95302 Cergy-Pontoise, France}

\begin{abstract}
	We investigate the dynamics of a particle in a confined periodic system---a time-dependent oscillator confined by infinitely high and moving walls---and focus on the evolution of the phase of the wavefunction. It is shown that, for some specific initial states in this potential, the phase of the wavefunction throughout the cavity depends on the walls motion. We further elaborate a thought experiment based on interferences devised to detect this form of single-particle nonlocality from a relative phase. We point out that, within the non-relativistic formalism based on the Schr\"odinger equation (SE), detecting this form of nonlocality can give rise to signaling. We believe this effect is an artifact, but the standard relativistic corrections to the SE do not appear to fix it. Specific illustrations are given, with analytical results in the adiabatic approximation, and numerical computations to show that contributions from high-energy states (corresponding to superluminal velocities) are negligible.

\end{abstract}
\maketitle
\section{Introduction}

While quantum nonlocality based on multi-particle entanglement is
well-recognized, single particle nonlocality remains controversial.
The primary candidate of the latter would be\footnote{We are interested here in the non-local aspects generated by the dynamics of the single particle wavefunction. This should not be conflated with single particle entanglement, involving usually spatially separated wavepackets of a photon having passed through a beamsplitter (see eg \cite{proc} for a discussion).} the Aharonov-Bohm effect
\citep{AharonovAB}. In the Aharonov-Bohm (AB) effect, the wavefunction
phase evolution is deemed to be nonlocal. The phase contains dynamical
and geometric components. Both the dynamical \citep{aharonov-tokyo,popescu-DNL}
and the geometric \citep{anandan88} components are ascribed a nonlocal
origin. Nevertheless, the nonlocal character of the Aharonov-Bohm
effect has been disputed on the ground that electromagnetic forces
might be able to account for the AB phase \citep{vaidman,batelaan}.
The nonlocal character of the quantum phase therefore remains controversial.

In this work, we investigate the same issue of phase nonlocality of
a single quantum particle but we focus on an entirely different system.
The system we will be dealing with is a time-dependent linear oscillator
confined by infinitely high walls, one of which is moving. Put differently,
our system is an infinite well subjected to a time-dependent harmonic
potential and in which one of the well's walls has an oscillatory
motion. The first reason for choosing this system is that analytic
solutions of the time-dependent Schr\"odinger equation are known \citep{makowski91}.
The second, less mundane reason, is that such systems, and in particular
their simplest variant (a box with a linearly moving wall) have long
been suspected of manifesting some form of nonlocality \citep{greenberger,makowski-p,wang2008,mmw18,colin20}.

It was indeed conjectured \citep{greenberger,makowski-p} that the
moving wall could nonlocally change the phase of a wavepacket at the
center of the box that remained localized far from the wall. While
this conjecture proved to be incorrect \citep{A2018}, it was recently
noted \citep{mmw18} that when a quantum state had a non-zero probability
amplitude near the wall, a linearly expanding wall induced instantaneously
a current density at any point of the box. We note for completeness
that systems with moving boundaries are of current interest in practical
schemes in the field of quantum engines or in atomic spectroscopy
\citep{elouard18,duffin18,photo18}.

In the present paper, we will be interested in the phase evolution
of a quantum state in a confined oscillator with a moving wall. More
precisely, we will focus on a particle in a cavity whose left wall
is fixed (say at $x=0$) but whose right wall oscillates, while inside
the cavity the particle is subjected to an oscillator potential. We
will require the initial state to be spread throughout the cavity
but the phase will only be measured near the static wall (at $x=0$)
of the cavity. For an appropriately chosen initial state $\psi(t_{0})$
and Hamiltonian parameters, the time evolved state after one period
of oscillation is simply $\psi(x,t_{0}+T)=e^{-i\mu}\psi(x,t_{0})$,
where $\mu$ is a global phase. The same initial state evolving in
an identical cavity but with a different motion for the moving wall
will not yield the same global phase, although the cause of the different
phase evolution is due to a potential that is different only in a
small region near the moving wall. 

Naturally these phases are not observable, but we elaborate a thought
experiment that allows an observer located near the fixed wall to
nonlocally infer the phase at time $t=T$ by splitting the state into
a spatial superposition at $t=0$ and observing the resulting interference
after the parts are recombined. Since the non-relativistic Schr\"odinger
equation does not impose an upper bound for energies or velocities,
the observer near the fixed wall can even infer the phase before a
light signal has the time to propagate from the moving wall. This
feature could give rise to some form of signaling (although our observer
needs to accumulate ensemble statistics in order to deduce the phase
difference and cannot infer anything in a single shot). We should
stress that we do not believe this signaling is physical, but rather
that it is an artifact of employing a non-relativistic formalism.
However, we will show that it is not clear how the tiny contributions
from high-energy eigenstates (with superluminal velocities) can account
for the observed change in the global phase.

We will start by introducing the Hamiltonian of the system and the
solutions to the Schr\"odinger equation (Sec. \ref{A}). We will then
focus on the phase evolution in particular by comparing the cases
in which the Hamiltonian is identical except at the moving boundary
(Sec. \ref{B}). In the latter case, we will give the solution in
the adiabatic approximation. Sec. \ref{C} deals first with the nonlocal
nature of the quantum phase, and then explains how the formalism can
result in signaling. A specific protocol will be given. We will then
illustrate our results by choosing a specific potential and boundary
motion (Sec. \ref{D}).

\section{Time-dependent oscillators with moving walls}

\label{A}

\subsection{Time-dependent linear oscillators}

A quantum time-dependent linear oscillator is a system comprising
a particle of mass $m$ subjected to the Hamiltonian, 
\begin{equation}
H_{TDLO}=\frac{P^{2}}{2m}+\frac{1}{2}m\Omega^{2}(t)x^{2},
\end{equation}
where we will assume in this work $\Omega^{2}(t)$ to be $T$-periodic,
$\Omega^{2}(t+T)=\Omega^{2}(t)$. The solutions $\phi(x,t)$ of the
Schr\"odinger equation, 
\begin{equation}
i\hbar\partial_{t}\phi(x,t)=H_{TDLO}\phi(x,t),
\end{equation}
depend on an initial condition $\phi(x,t=0)=\phi_{0}(x)$ and on appropriate
boundary conditions, eg $\phi(x\rightarrow\pm\infty,t)\rightarrow0$.
These solutions can be obtained in semi-analytical form. Several methods
have been developed, from the one relying on obtaining the eigenfunctions
of dynamical invariants (see \citep{pedrosa2008} and Refs therein),
pioneered by Lewis and Riesenfeld \citep{LR}, to more general Lie
system based approaches \citep{carinena2011}.

A time-dependent linear oscillator can be confined in a box bounded
by infinitely high fixed walls (FW). Let $L_{0}$ denote the width of the well,
ie the distance between the two walls.\ The corresponding Hamiltonian
is given by, 
\begin{align}
H_{FW} & =\frac{P^{2}}{2m}+V\label{hamFW}\\
V(x) & =\left\{ \begin{array}{l}
\frac{1}{2}m\Omega^{2}(t)x^{2}\text{ \ for}\ \ 0\leq x\leq L_{0}\\
+\infty\text{ \ otherwise}
\end{array}\right.,\label{vdef}
\end{align}
and the solutions $\phi(x,t)$ of the Schr\"odinger equation must respect
the boundary conditions $\phi(x=0,t)=\phi(x=L_{0},t)=0$. There are
no general methods\ to solve the confined time-dependent (nor the
confined standard harmonic) oscillator (see \citep{razavy,ghosh}
and Refs therein for specific cases).

\subsection{Confined oscillator with time-dependent boundary conditions\label{mwt}}

\begin{figure}
\centering \includegraphics[width=5in]{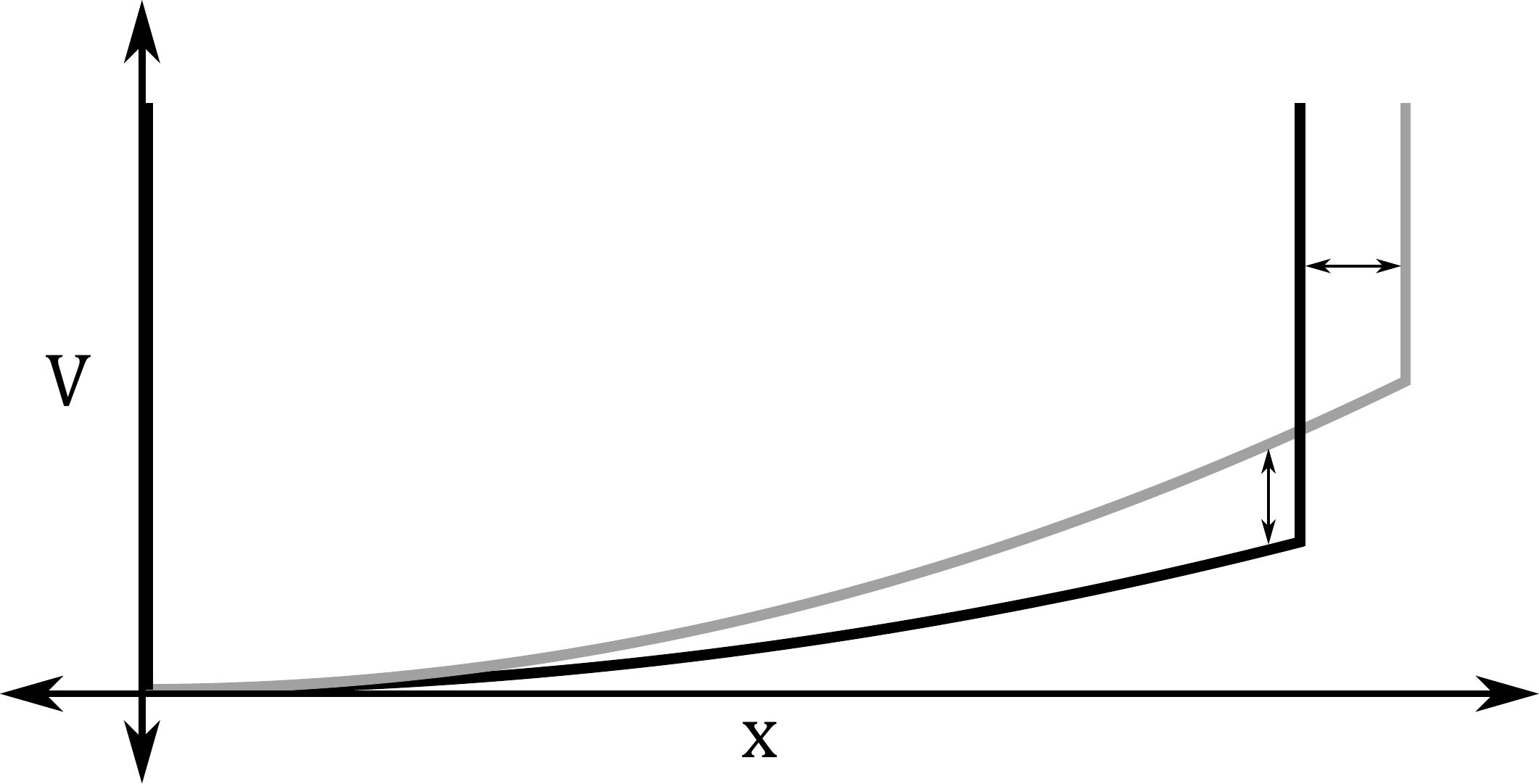} \caption{Confined harmonic oscillator with the harmonic frequency $\Omega(t)$
and the position of the left wall $L(t)$ periodic functions of time. }
\label{HWell} 
\end{figure}

We now confine the time-dependent oscillator in a box with a moving
outer wall, i.e. the box has length \ $L(t),$ with boundaries at
$x=0$ and $x=L(t)$, as shown in Fig. \ref{HWell}. The Hamiltonian
is similar to Eqs. (\ref{hamFW})-(\ref{vdef}): 
\begin{align}
H & =\frac{P^{2}}{2m}+V\label{ham}\\
V(x) & =\left\{ \begin{array}{l}
\frac{1}{2}m\Omega^{2}(t)x^{2}\text{ \ for}\ \ 0\leq x\leq L(t)\\
+\infty\text{ \ otherwise}
\end{array}\right.\label{pot}
\end{align}
This system is mathematically intricate as it involves a different
Hilbert space at each time $t$ and standard quantities such as the
the time derivative are ill-defined. Nevertheless, solutions of the
Schr\"odinger equation for specific boundary conditions $L(t)$ and
related frequencies $\Omega_{L}(t)$ are known \citep{makowski91}.
While a proper \citep{mosta1999,martino} approach involves using time-dependent
unitary transformations mapping the time-dependent boundary conditions
to fixed ones, it is more straightforward to verify by direct substitution
that, 
\begin{equation}
\psi_{n}(x,t)=\sqrt{2/L(t)}\exp\left(-i\frac{\hbar\pi^{2}n^{2}}{2m}\int_{t_{0}}^{t}\frac{1}{L^{2}(t^{\prime})}dt^{\prime}\right)\exp\left(i\frac{m}{2\hbar}\frac{\partial_{t}L(t)}{L(t)}x^{2}\right)\sin\frac{n\pi x}{L(t)},\label{eig}
\end{equation}
obeys the Schr\"odinger equation, 
\begin{equation}
i\hbar\partial_{t}\psi_{n}(x,t)=\frac{-\hbar^{2}}{2m}\partial_{x}^{2}\psi_{n}(x,t)+\frac{1}{2}m\Omega^{2}(t)x^{2}\psi_{n}(x,t),
\end{equation}
with the boundary conditions, 
\begin{equation}
\psi_{n}(0,t)=\psi_{n}(L(t),t)=0,
\end{equation}
provided that, 
\begin{equation}
\Omega^{2}(t)=-\frac{\partial_{t}^{2}L(t)}{L(t)}.\label{freq}
\end{equation}

The $\psi_{n}(x,t)$ are not instantaneous eigenstates, but form a
set of orthogonal basis functions that can be used to determine the
evolution of an arbitrary initial state. We will choose in this work
$L(t)$ and therefore $\Omega^{2}(t)$ to be $T-$periodic functions.
An important property is the phase increment after a full period.
Indeed, comparing $\psi_{n}(x,t_{0})$ and $\psi_{n}(x,t_{0}+T)$
leads immediately to, 
\begin{equation}
\psi_{n}(x,t_{0}+T)=e^{-i\mu_{n}}\psi_{n}(x,t_{0}),\label{fazeq}
\end{equation}
with, 
\begin{equation}
\mu_{n}=\frac{\hbar\pi^{2}n^{2}}{2m}\int_{t_{0}}^{t_{0}+T}\frac{1}{L^{2}(t^{\prime})}dt^{\prime}.\label{mudef}
\end{equation}
Hence after a full cycle, a time-dependent oscillator with moving
walls in state $\psi_{n}$ returns to its initial state except for
a phase increment $\mu_{n}$$.$ Following Aharonov and Anandan \citep{aharonov-anandan},
the total phase $\mu_{n}$ can be parsed into a ``dynamical''\ part
$\delta_{n}$ encapsulating the usual phase increment due to the instantaneous
expectation value of the Hamiltonian and a ``geometric''\ part
$\gamma_{n}$. The dynamical phase, 
\begin{equation}
\delta_{n}=\hbar^{-1}\int_{t_{0}}^{t_{0}+T}\left\langle \psi_{n}(t^{\prime})\right\vert H\left\vert \psi_{n}(t^{\prime})\right\rangle dt^{\prime},
\end{equation}
is readily computed \citep{A2018} and the nonadiabatic geometric
phase, $\gamma_{n}$ is then obtained as, 
\begin{equation}
\gamma_{n}=\mu_{n}-\delta_{n}=\frac{m}{12\hbar\pi^{2}n^{2}}\left(2\pi^{2}n^{2}-3\right)\int_{t_{0}}^{t_{0}+T}\left[L(t)\partial_{t}^{2}L(t)-\left(\partial_{t}L(t)\right)^{2}\right]dt.\label{faz}
\end{equation}
Note that in the case of a free confined Hamiltonian, it is possible to obtain a general expression for the geometric phase associated to a cyclic adiabatic evolution in terms of the chosen boundary conditions \cite{facchi}.

\section{The wavefunction phase evolution}

\label{B}

\subsection{Phase and walls motions \label{cases}}

The phase $\mu_{n}$ {[}Eq. (\ref{mudef}){]} is a property of the
\emph{entire} wavefunction, although part of the phase increment is
due to the walls' motion. To see this, we will compare two oscillators
that have exactly the same potential everywhere except in the vicinity
of the outer wall. To this end we will examine three different cases: 
\begin{enumerate}
\item an oscillator with the outer wall moving according to some function
$L_{1}(t)$ and with a time-dependent frequency obeying Eq. (\ref{freq}),
$\Omega_{1}^{2}(t)=-\partial_{t}^{2}L_{1}(t)/L_{1}(t);$ 
\item an oscillator with the outer wall moving according to a function $L_{2}(t)$
and with a time-dependent frequency also obeying Eq. (\ref{freq}),
$\Omega_{2}^{2}(t)=-\partial_{t}^{2}L_{2}(t)/L_{2}(t);$ 
\item an oscillator with the outer wall moving according to the function
$L_{2}(t)$ but evolving in the potential of the first case, $\Omega_{1}^{2}(t)=-\partial_{t}^{2}L_{1}(t)/L_{1}(t)$. 
\end{enumerate}
We will take $L_{2}(t)$ to be very close to $L_{1}(t)$ and with
the same period $T$. The idea is to compare the phase in cases 1
and 3, which evolve in the same potential $\Omega_{1}(t)$ except
in the region near the walls, since the boundary conditions, depending
respectively on $L_{1}(t)$ and $L_{2}(t)$, are slightly different.
We will start from the same initial state $\psi(x,t=0)$ of the type
given by Eq.(\ref{eig}). In order to have the same initial state,
we must impose $L_{1}(t=0)=L_{2}(t=0)$ and $\partial_{t}L_{1}(t=0)=\partial_{t}L_{2}(t=0).$
Picking functions for which $\partial_{t}L_{1}(t=0)=\partial_{t}L_{2}(t=0)=0$
leads to an initial state, 
\begin{equation}
\psi(x,t=0)=\sqrt{2/L_{0}}\sin\frac{n\pi x}{L_{0}},
\end{equation}
with $L_{0}=$ $L_{1}(t=0)=L_{2}(t=0)$.

The time-evolved wavefunction in cases I and II is obtained directly
from Eq. (\ref{eig}). The total phase increment after one period
$\mu_{n}^{\text{I}}$ and $\mu_{n}^{\text{II}}$ is given by Eq. (\ref{mudef}).\ Case
III\ however does not respect Eq. (\ref{freq}) and therefore does
not fit in the framework developed in Sec. \ref{mwt}. We will look
for a perturbative solution in this case. Note that in case III there
is no reason to expect the existence of a global phase increment,
though due to the continuity of the wavefunction, we can expect that
if $L_{1}(t)$ and $L_{2}(t)$ are close enough, the phase increment
after one full cycle will not deviate far from a constant value in
the vicinity of a given point. We will be interested in the phase
in the vicinity of $x=0$, which is the farthest region from the moving
walls.

\subsection{Phase increment in case \emph{III \label{subsec:B2}}}

\subsubsection{Quantum state evolution}

Let $\phi(x,t)$ denote the solution
of the Schr\"odinger equation for the case III mentioned in the preceding
subsection. Let $\psi_{n}^{\text{II}}(x,t)$ denote the basis functions
of Eq. (\ref{eig}) with $L(t)=L_{2}(t).$ Since $\phi(x,t)$ obeys
the boundary condition $\phi(L_{2}(t),t)=0$, we will look for a solution
in the form, 
\begin{equation}
\phi(x,t)=\sum_{k=1}^{\infty}a_{k}(t)\psi_{k}^{\text{II}}(x,t),\label{exp}
\end{equation}
with the initial condition being, 
\begin{equation}
\phi(x,0)=\psi_{n}^{\text{II}}(x,0)\text{ \ or }a_{k}(0)=\delta_{kn}\text{\ }.\label{coeffci}
\end{equation}

Applying the Schr\"odinger equation with $V=\frac{1}{2}m\Omega_{1}^{2}(t)x^{2}$
for $0<x<L_{2}(t)$ to the right hand-side of Eq. (\ref{exp}) and
recalling that $i\hbar\partial_{t}\psi_{k}^{\text{II}}(x,t)=\left(\frac{P^{2}}{2m}+\frac{1}{2}m\Omega_{2}^{2}(t)x^{2}\right)\psi_{k}^{\text{II}}(x,t)$
leads to, 
\begin{equation}
\sum_{k}i\hbar\partial_{t}a_{k}(t)\psi_{k}^{\text{II}}(x,t)=\sum_{k}\frac{1}{2}m\left(\Omega_{1}^{2}(t)-\Omega_{2}^{2}(t)\right)x^{2}a_{k}(t)\psi_{k}^{\text{II}}(x,t),\label{set}
\end{equation}
and by projecting $\left\langle \psi_{j}^{2}\right\vert $ we obtain,
\begin{equation}
\partial_{t}a_{j}(t)=\frac{-i}{\hbar}\sum_{k}a_{k}(t)\frac{1}{2}m\left(\Omega_{1}^{2}(t)-\Omega_{2}^{2}(t)\right)\int_{0}^{L_{2}(t)}x^{2}\left[\psi_{j}^{2}(x,t)\right]^{\ast}\psi_{k}^{\text{II}}(x,t)dx.\label{coefft}
\end{equation}
The integral can be readily computed, yielding, 
\begin{equation}
\frac{8kj(-1)^{k+j}L_{2}(t)^{2}\exp\left(-i\hbar\pi^{2}(k^{2}-j^{2})\int L_{2}(t')^{-2}dt'/\left(2m\right)\right)}{\pi^{2}\left(k^{2}-j^{2}\right)^{2}},\label{rct}
\end{equation}
for $k=j,$the integral becomes, 
\begin{equation}
\frac{\left(2j^{2}\pi^{2}-3\right)L_{2}(t)^{2}}{6j^{2}\pi^{2}}.
\end{equation}

In the \emph{adiabatic} case, only the coefficient of the initial
state is non-zero, $a_{j}(t)=a_{k}(t)=\delta_{nk}\delta_{nj}a_{n}(t)$.
There is no summation in Eq. (\ref{coefft}) and $a_{n}(t)$ is obtained
by solving analytically the differential equation (see  Eq. (\ref{aad}) below).
The integral in that equation may be obtained analytically for specific
choices of $L_{1}(t)$ and $L_{2}(t)$; otherwise a simple numerical
integration is in order.

In the \emph{generic} (nonadiabatic) case there are no analytical
solutions. We obtain numerical solutions by writing first Eq. (\ref{set})
as 
\begin{equation}
\sum_{k=1}^{\infty}i\hbar\partial_{t}\left[a_{k}(t)\psi_{k}^{\text{II}}(x,t)\right]=\sum_{k=1}^{\infty}a_{k}(t)\left[\frac{-\hbar^{2}}{2m}\partial_{x}^{2}\psi_{k}^{\text{II}}(x,t)+\frac{1}{2}m\Omega_{1}^{2}(t)x^{2}\psi_{k}^{\text{II}}(x,t)\right],\label{ub}
\end{equation}
and then multiplying this equation by $\left(\psi_{j}^{\text{II}}(x,t)\right)^{*}$
and integrating. The basis functions are orthogonal, and the remaining
non-trivial integrals 
\begin{equation}
\int_{0}^{L_{2}(t)}\left[\psi_{j}^{\text{II}}(x,t)\right]^{*}\partial_{t}\psi_{k}^{\text{II}}(x,t)dx\text{ and \ensuremath{\int_{0}^{L_{2}(t)}\left[\psi_{j}^{\text{II}}(x,t)\right]^{*}x^{2}\psi_{k}^{\text{II}}(x,t)dx}}
\end{equation}
can be integrated analytically without difficulty. This leaves us
with a system of linear first-order (in time) coupled equations for
$a_{k}(t)$, subject to the initial condition $a_{k}(t=0)=\delta_{nk}.$
This formally infinite system is truncated by setting an upper bound
$k_{max}$ in the sum of Eq. (\ref{ub}). The choice of $k_{max}$
depends on the desired accuracy of the solutions: since $\left|a_{k}(t)\right|\rightarrow0$
as $k$ increases, it should be checked that $\left|a_{k_{max}}(t)\psi_{k_{max}}^{\text{II}}(x,t)\right|<z$
throughout the $x$ and $t$ intervals over which the solutions are
computed ($z$ is the numerical zero). 

Note that the dynamical phase for case III can be obtained by computing,

\begin{align}
\delta_{n} & =\hbar^{-1}\int_{0}^{T}\left\langle \psi_{n}^{\text{II}}(t^{\prime})\right\vert \left(\frac{P^{2}}{2m}+\frac{m}{2}\Omega_{1}^{2}(t)X^{2}\right)\left\vert \psi_{n}^{\text{II}}(t^{\prime})\right\rangle dt^{\prime}\\
 & =\frac{\hbar\pi^{2}n^{2}}{2m}\int_{0}^{T}\frac{1}{L_{2}^{2}(t^{\prime})}dt^{\prime}+\frac{m}{12n^{2}\pi^{2}\hbar}\left(2\pi^{2}n^{2}-3\right)\int_{0}^{t}\left(\left(\partial_{t^{\prime}}L_{2}(t^{\prime})\right)^{2}-\frac{\partial_{t^{\prime}}^{2}L_{1}(t^{\prime})}{L_{1}(t^{\prime})}L_{2}^{2}(t^{\prime})\right).\label{dynfazmod}
\end{align}

\subsubsection{Adiabatic approximation} To obtain the coefficients $a_{k}(t)$
in closed form, approximations need to be made. The simplest regime
is the ``adiabatic''\ approximation which involves neglecting the
contribution of basis states other then the initial one, as specified
by the initial condition (\ref{coeffci}). Hence $a_{k}(t)=\delta_{nk}a_{n}(t)$
and $a_{n}(t)$ is obtained from Eq. (\ref{coefft})
by keeping only the diagonal contribution, yielding, 
\begin{equation}
a_{k}(t)=\delta_{nk}\exp\left(-\frac{i}{\hbar}\frac{m}{12n^{2}\pi^{2}}\left(2\pi^{2}n^{2}-3\right)\int_{0}^{t}\left(L_{2}(t^{\prime})\partial_{t^{\prime}}^{2}L_{2}(t^{\prime})-\frac{\partial_{t^{\prime}}^{2}L_{1}(t^{\prime})}{L_{1}(t^{\prime})}L_{2}^{2}(t^{\prime})\right)dt^{\prime}\right).\label{aad}
\end{equation}
Note that the adiabatic approximation employed here is similar but not strictly equivalent to the standard adiabatic theorem (see e.g. Ch. 2 of \cite{bookGP}), since the adiabatic theorem deals with instantaneous eigenstates of a single Hamiltonian, whereas here we are employing basis functions of the case II Hamiltonian in the Schr\"odinger equation with the case III Hamiltonian.
Plugging this back this expresiion for the $a_{k}(t)$ coefficients into Eq. (\ref{exp}), we see that within our
approximation the total phase after one full cycle is, 
\begin{equation}
\mu_{n}^{ad}=\frac{\hbar\pi^{2}n^{2}}{2m}\int_{0}^{T}\frac{1}{L_{2}^{2}(t^{\prime})}dt^{\prime}+\frac{m}{12n^{2}\pi^{2}\hbar}\left(2\pi^{2}n^{2}-3\right)\int_{0}^{T}\left(L_{2}(t^{\prime})\partial_{t^{\prime}}^{2}L_{2}(t^{\prime})-\frac{\partial_{t^{\prime}}^{2}L_{1}(t^{\prime})}{L_{1}(t^{\prime})}L_{2}^{2}(t^{\prime})\right)dt^{\prime}.\label{muad}
\end{equation}
The first term is the case II phase $\mu_{n}^{\text{II}}$, and the
second term appears as a correction. Part of this correction is due
to the dynamical phase, which is different from case II since the
potential in case III is the one from case I, leading to the dynamical
phase given in Eq.(\ref{dynfazmod}). The remaining part of this correction,
proportional to $\int_{0}^{T}\left(L_{2}(t^{\prime})\partial_{t^{\prime}}^{2}L_{2}(t^{\prime})-\left(\partial_{t^{\prime}}L_{2}(t^{\prime})\right)^{2}\right)dt^{\prime}$
is a geometric term due to the boundary conditions.

The adiabatic approximation is expected to hold when $L_{1}(t)$ and
$L_{2}(t)$ are almost identical. Then, in order to discriminate the
phase of the different cases mentioned above, it is crucial that $\mu_{n}^{\text{I}}$
and $\mu_{n}^{\text{II}}$ differ significantly. Indeed, $\mu_{n}^{ad}$
appears as a correction to\ $\mu_{n}^{\text{II}}$ {[}Eq. (\ref{muad}){]},
so that ensuring that $\mu_{n}^{ad}\approx\mu_{n}^{\text{II}}\neq\mu_{n}^{\text{I}}$
while still having $L_{1}(t)\approx L_{2}(t)$ typically implies high
values of $n$ and/or small values of $m$ {[}see Eq. (\ref{mudef}){]}.\\

\subsubsection{Generic case} In general, the adiabatic approximation will
of course not be valid. The generic case is characterized by a set
of non-negligible coefficients $a_{k}(t)$ with $k$ lying in the
interval $[n-N$,$n+N]$. Indeed, initially only $a_{n}(t=0)$ is
non-vanishing, and the coupling between $a_{n}(t=0)$ and the different
coefficients $a_{k}(t)$ falls off as $1/k^{3}$ for large $k$ (see
Eqs. (\ref{coefft}) and (\ref{rct})). Since
here we are interested in short time evolutions, we can expect that
only a few basis states $\psi_{k}$ centered on $k=n$ will contribute
in the expansion Eq. (\ref{exp}). While there is no simple analytical
formula giving the phase increment in this generic situation, it is
straightforward to compute numerically the wavefunction in case III
and from there extract the phase increment after one full cycle. Note
that in general there is no reason to expect that similarly to Eq.
(\ref{fazeq}), the wavefunction after one period will be equal to
the initial wavefunction up to a global phase. We can however expect,
for reasonable choices of $L_{1}(t)$ and $L_{2}(t),$ the phase to
be slowly varying (as a function of $x$) and approximately constant
in the neighborhood of $x=0$, which will be our region of interest
in the protocols described below.

\section{Nonlocality and signaling}

\label{C}

\subsection{Nonlocal origin of the quantum phase}

Let us go back to the three cases described in Sec.\ \ref{cases},
from the point of view of an observer placed in the neighborhood of
$x=0$. Let the system be initially prepared in a state $\psi_{n}(t=0)$
given by Eq.\ (\ref{eig}). Since we want $\psi_{n}(t=0)$ to be
independent of whether the system will evolve with boundary functions
$L_{1}(t)$ or $L_{2}(t)$, we must enforce $L_{1}(t=0)=L_{2}(t=0)$
and for simplicity we set $\partial_{t}L_{1}(t=0)=\partial_{t}L_{2}(t=0)=0$.
Let us assume the observer, say Alice, can determine the phase difference
between the initially prepared state $\psi_{n}(t=0)$ and the state
after one period $\psi_{n}(t=T)$. Alice can therefore discriminate
case I from case II, since the phases at $t=T,$ $\mu_{n}^{\text{I}}$
and $\mu_{n}^{\text{II}}$ {[}Eq. (\ref{mudef}){]} will be different.
But the Hamiltonian is also different in these two cases, a point
Alice could have checked by making local measurements in her vicinity,
so she won't be surprised by finding different phases depending on
whether the system evolved in case I\ or case II.

However, when comparing cases I\ and III, the Hamiltonian is identical
except in the vicinity of the wall's position: only the wall's motion
differs in cases I and III.\ Nevertheless, the total phase at $t=T$
will be different, including in the region where Alice is standing.\ Hence
the phase difference between cases I\ and III must be attributed
to the potential in the region near the opposite wall. Since the wall
can be arbitrarily far from the $x=0$ region, we can say that the
phase difference appears to be due to local potentials varying in
an arbitrarily remote region: the phase increment, as it appears in
the region near the static wall at $x=0$, has a nonlocal origin.

\subsection{Signaling}

Let us still assume that Alice has access to \ the phase difference
between the wavefunctions at $t=0$ and $t=T$ (such a protocol is
given immediately below), and further assume that the walls are sufficiently
far away so that the time it takes for a light signal emitted from
the moving (right) wall to reach her position near the fixed left
wall, $\tau\simeq L(t=0)/c$ is larger than the period $T$ of the
wall's motion. By measuring the phase difference, she can determine
whether the wall is moving according to $L_{1}(t)$ or $L_{2}(t).$
Alice can thus discriminate case I from case III before a signal sent
from the moving wall, say by Bob, reaches her. In principle, by choosing
different functions $L_{2}(t)$, Bob could send signals to Alice superluminally.

\begin{figure}[tb]
\includegraphics[width=5in]{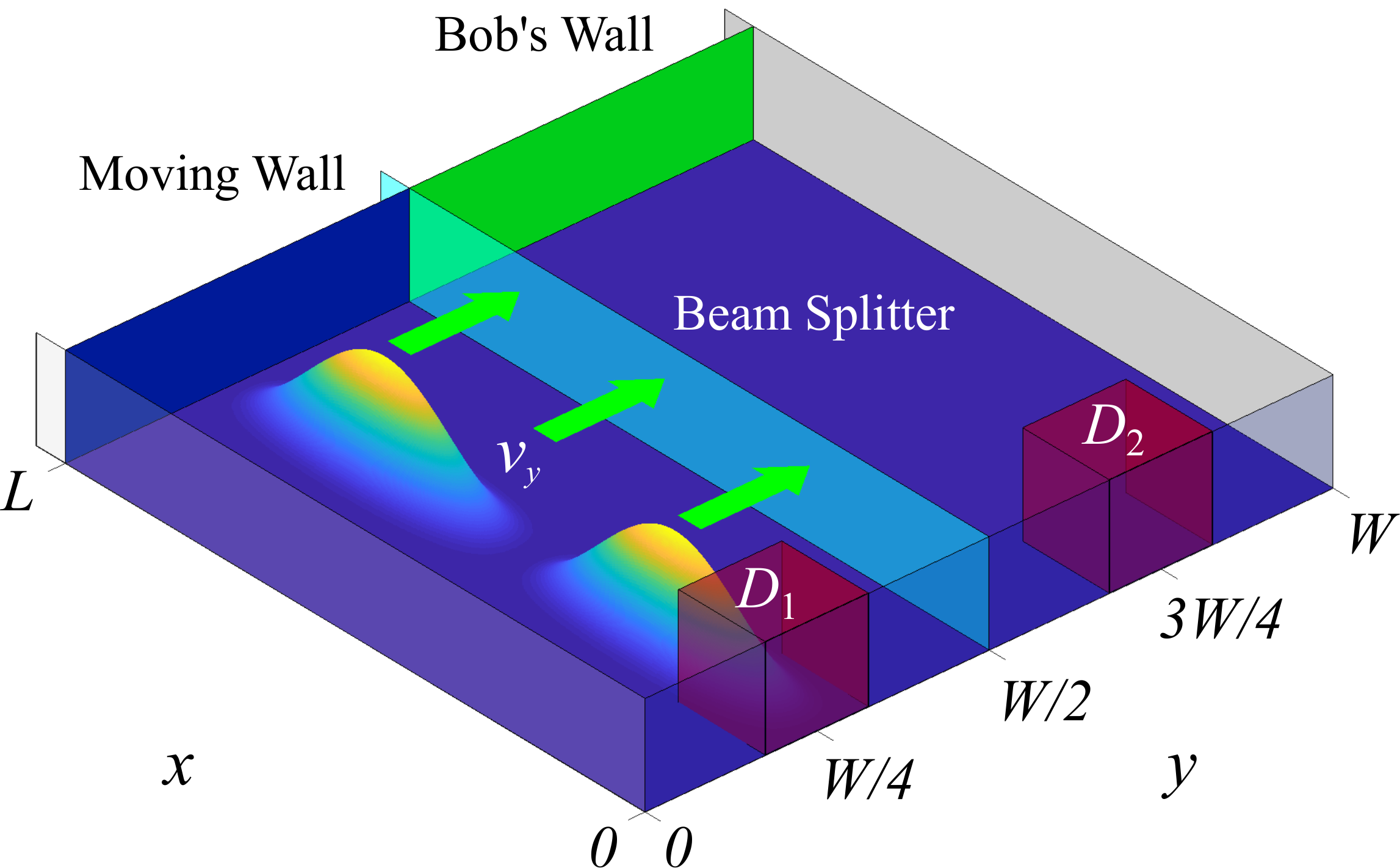} \caption{A thought experiment allowing communication via the nonlocal phase
of the wavefunction. The walls of the cavity are initially at rest,
the harmonic potential is turned off, and the detectors are not yet
in place. The particle begins in the state $\psi_{n}(x,t=0)g(y,t=0)$,
where $\psi_{n}(x,=0)$ is an excited $x$-mode of the cavity ($n=2$
is depicted), and $g(y,t=0)$ is a narrow Gaussian initially centered
at $W/4$ with average velocity $v_{y}>0$. When the packet has divided
at the beam splitter, the harmonic potential $(1/2)m\Omega_{1}^{2}(t)x^{2}$
is turned on throughout the cavity, and the wall segment at $(x=L,y<W/2)$
begins to move as $L_{1}(t)$. At the same moment, Bob chooses whether
the wall segment at $(x=L,y>W/2)$ begins to move as $L_{1}(t)$ (message
0) or $L_{2}(t)$ (message 1). The two half-packets propagate along
$y$, then bounce off their respective walls and meet back at the
beam splitter, a period $T=W/v_{y}$ after they left it. Bob choosing
$L_{1}(t)$ results in perfect interference, so the entire pulse recombines
on the $y>W/2$ side of the beam splitter, whereas choosing $L_{2}(t)$
results in a phase difference so that the interference is no longer
perfect. Alice can detect Bob's choice at $t=3T/2=3W/2v_{y}$ by inserting
detector $D_{1}$ at $W/4$ and detector $D_{2}$ at $3W/4$ and measuring
the relative intensity for the entire ensemble of cavities. $L_{1}(t)$,
$L_{2}(t)$, and $\Omega_{1}(t)$ are all periodic function of time,
with period $T$.}
\label{figFTL} 
\end{figure}

\subsection{Protocol}

We describe here one possible protocol that formally leads to signaling.
Indeed, by changing the motion of the wall at $x=L$, Bob is able
to change the global phase evolution, which is an instantaneous and
measurable effect that occurs everywhere in the cavity. Because of
this, Bob can send a message by choosing how the wall moves after
$t=0$, and then Alice can detect this choice by making local measurements
near $x=0$. To make sure the message is clearly resolved, Alice and
Bob need to share a large ensemble of identical cavities, and Bob
must make the same choice of motion for all of them. Provided that
Alice completes her measurements well before $t=\tau\simeq L(0)/c$,
this message is sent faster than $c$.

In order for Alice to perform her measurement, the cavity will need
to have a bit more structure (see Fig. \ref{figFTL}). Suppose that
in the $y$-direction, the cavity has the potential of a fixed infinite
square well of width $W\ll L$, with one wall at $y=0$, the other
at $y=W$, and a beam splitter at $y=W/2$, which runs the full length
from $x=0$ to $x=L$. The harmonic potential is initially turned
off and the walls are stationary.

The initial state of the particle is $\psi(x,t=0)g(y,t=0)$, where
$\psi(x,t=0)$ is the cavity state we have been discussing, and $g(y,t=0)$
is a Gaussian wave-packet with average $y$-velocity $v_{y}>0$ which
begins well-localized at $y=W/4$, as displayed in Fig. \ref{figFTL}.
We assume that this experiment will finish quickly enough that the
spreading of this wave-packet can be ignored.

When the particle strikes the beam splitter, half of the wavefunction
is reflected and half is transmitted. At the moment the packets have
passed the beam splitter, the harmonic potential $\Omega_{1}(t)$
is turned on throughout the cavity, and the wall segment at $x=L$
and $y<W/2$ begins to move according to $L_{1}(t)$. Bob also chooses
at this moment whether the wall segment at $x=L$ and $y>W/2$ begins
to move according to $L_{1}(t)$ (message 0) or $L_{2}(t)$ (message
1). The two half-packets propagate along $y$ then bounce off their
respective walls, and meet back at the beam splitter, a period $T=W/v_{y}$
after they left it --- which is incidentally when $L_{1}(t)=L_{2}(t)$
again. The harmonic potential and wall motion are then turned off
once more.

The cavity is tuned so that if Bob chose $L_{1}(t)$, then when the
two half-pulses meet, they interfere destructively for $y<W/2$ and
constructively for $y\geq W/2$, and thus the particle always ends
up in the region $y\geq W/2$. However, if Bob chose $L_{2}(t)$,
then the two half-pulses would have accumulated a phase difference
of $\Delta \mu$ before they meet again, and the interference would no longer
be perfectly constructive/destructive.

At $t=3T/2=3W/2v_{y}$, Alice places detector $D_{1}$ at $y=W/4$
and detector $D_{2}$ at $y=3W/4$. Each detector spans $0<x<\epsilon$,
and the entire effective width of the Gaussian packet in the $y$-direction.
If Bob chose the wall motion $L_{1}(t)$, only $D_{2}$ can fire since
the particle is in the upper half of the cavity, but since the detectors
have a small width $\epsilon$ relative to the cavity length $L$,
the probability that Alice detects a particle in each cavity is only
approximately $(2n^{2}\pi^{2}/3)(\epsilon/L)^{3}$. Hence Alice and
Bob need to share a large ensemble of identical cavities as this allows
Alice to count the total number of clicks on the detectors $D_{1}$
and $D_{2}$. In particular, if Bob has chosen $L_{1}(t)$ the phase
generated by the wall motion will lead to destructive interference
in the lower part of the cavity, so Alice will only obtain clicks
on $D_{2}$, and thus $\langle D_{1}\rangle/\langle D_{2}\rangle=0$
indicating message 0 (where $\langle D_i\rangle$ is the average number of detector clicks per cavity); instead, if Bob's wall moves as $L_{2}(t)$,
then this generates a phase difference $\Delta \mu$ that leads to a phase difference in both cavities, and thus $\langle D_{1}\rangle/\langle D_{2}\rangle=\tan^{2}(\Delta\mu/2)$
indicating message 1 (an example derivation of $\Delta\mu$ can be found
in Sec. \ref{sec:Ill}). As a result, provided $3W/2L\ll v_{y}/c$
and assuming that counting the clicks takes a negligible time, Alice
is able to receive Bob's message before a signal traveling at $c$
could reach her.

The faster-than-light communication protocol can be summarized as
follows: 
\begin{itemize}
\item A large ensemble of quantum particles (electrons, say) are identically
prepared at $t=0$ in identical cavities as shown in Fig. \ref{figFTL}. 
\item At $t=T/4$ Bob chooses to send message 0 by setting the motion of his wall in every cavity of the ensemble to $L_{1}(t)$, or message 1 by setting their motion to
$L_{2}(t)$. 
\item At $t=3T/2$, Alice places detector $D_{1}$ at $y=W/4$ and detector
$D_{2}$ at $y=3W/4$ in every cavity of the ensemble. The ratio $\langle D_{1}\rangle/\langle D_{2}\rangle=0$
indicates message 0, and $\langle D_{1}\rangle/\langle D_{2}\rangle=\tan^{2}(\Delta\mu/2)$
indicates message 1. 
\end{itemize}

\section{Illustration}

\label{sec:Ill}

\label{D}

\subsection{System Hamiltonian}

For the purpose of illustration, let us choose the following wall
motion functions, 
\begin{equation}
L_{j}(t)=L_{0}+q_{j}(\cos\omega t-1),\label{moti}
\end{equation}
where $L_{0}=L(t=0)$ and $q_{j}\,\ll L_{0}$. Both walls move according
to Eq. \ref{moti}, but with the different amplitudes, $q_{j}$. Index
$j=1$ is case I from section \ref{cases}, with $\Omega_{1}^{2}(t)=-\partial_{t}^{2}L_{1}(t)/L_{1}(t)$,
and $j=2$ is case III, with the same $\Omega_{2}^{2}(t)=\Omega_{1}^{2}(t)$
and an independent $L_{2}(t)$. The Hamiltonian for cases I\ and
III is obtained, from Eqs. (\ref{ham})-(\ref{pot}) and (\ref{moti})
as, 
\begin{eqnarray}
H_{j} & = & \frac{P^{2}}{2m}+V_{j}\label{hamiex}\\
V_{j}(x) & = & \left\{ \begin{array}{l}
\frac{1}{2}m\frac{q_{1}\omega^{2}\cos\omega t}{L_{0}+q_{1}\left(\cos\omega t-1\right)}x^{2}\text{ \ for}\ \ 0\leq x\leq L_{j}(t)\\
+\infty\text{ \ otherwise}
\end{array}\right.,\label{hamiex2}
\end{eqnarray}
that is the potential differs only in the interval between $L_{1}(t)$
and $L_{2}(t)$.

In case I, the phase increment $\mu_{n}^{\text{I}}$ is calculated
from Eq. (\ref{mudef}) as, 
\begin{equation}
\mu_{n}^{\text{I}}=\frac{\hbar\pi^{3}n^{2}(L_{0}-q_{1})}{m\omega\left(L_{0}(L_{0}-2q_{1})\right)^{3/2}}.\label{muil1}
\end{equation}
In case III the phase increment can be obtained in the adiabatic approximation
from Eq. (\ref{muad}), giving explicitly, 
\begin{equation}
\mu_{n}^{ad}=\frac{\hbar\pi^{3}n^{2}(L_{0}-q_{2})}{m\omega\left(L_{0}(L_{0}-2q_{2})\right)^{3/2}}+\frac{m}{\hbar}\frac{1}{12n^{2}\pi^{2}}\left(2\pi^{2}n^{2}-3\right)\frac{2\pi\omega L_{0}^{2}\left(\sqrt{L_{0}(L_{0}-2q_{1})}-L_{0}+q_{1}\right)(q_{1}-q_{2})^{2}}{q_{1}^{2}\sqrt{L_{0}(L_{0}-2q_{1})}}.\label{muill}
\end{equation}
The adiabatic approximation requires here $q_{2}=q_{1}+\varepsilon$,
with $\varepsilon\ll q_{1}$, and $L_{0}\gg q_{1}$,$q_{2}$.

\subsection{Examples displaying nonlocality}

We now give a couple of numerical examples based on the Hamiltonians
(\ref{hamiex})-(\ref{hamiex2}) with parameters giving rise to nonlocality.\ As
stated above, this is defined when the period, now $T=2\pi/\omega$
is smaller than the time it takes a light signal to travel from the
initial moving wall's position to the static leftward wall, $\tau=L_{0}/c.$
The numerical computations for the case III wavefunctions $\phi(x,t)$
are carried out by solving the truncated version of the coupled system
defined by Eqs. (\ref{set})-(\ref{coefft}). This is similar to numerical
methods used in previous related works \citep{glasser,mmw18} except
that the expansion basis is taken to be the solutions of case II rather
than the instantaneous eigenstates (ie, those obeying $H\xi_{n}(x,t)=E_{n}(t)\xi_{n}(x,t)$).

\begin{figure}[tb]
\includegraphics[width=7cm]{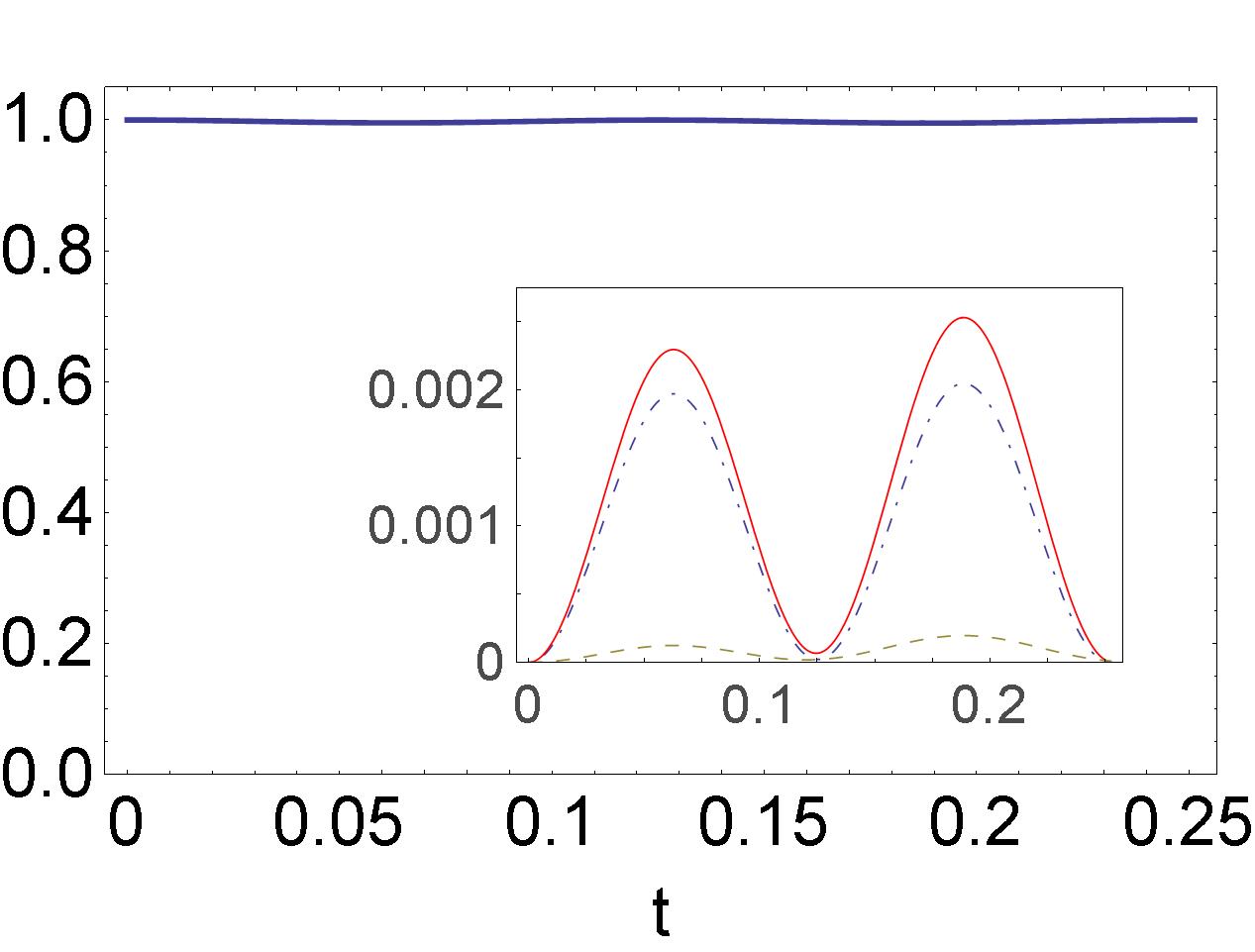}

\caption{Time evolution of the coefficients $a_{k}(t)$ of Eq. (\ref{exp})
in the \emph{adiabatic} case, corresponding to the Hamiltonian given
by Eqs. (\ref{hamiex})-(\ref{hamiex2}) with the following parameters
(in natural atomic units): $m=1/75,L_{0}=37,q_{1}=7,q_{2}=7.04,\omega=25$
(verifying $\left(T=2\pi/\omega\right)<\left(\tau=L_{0}/c\right)$
where $c$ is the light velocity). The initial state is $\phi(x,0)=\psi_{n=2}^{\text{II}}(x,0)$.
The thick solid blue line represents the numerically computed values
of $\left\vert a_{2}(t)\right\vert ^{2}$ while the inset shows the
corrections to the adiabatic approximation stemming from exact numerical
computations: $\left\vert a_{k}(t)\right\vert ^{2}$ for $k=1,3,4$
are represented by the blue dot-dashed, solid red and dashed green
curves, respectively.}
\label{figex1} 
\end{figure}

\begin{figure}[tb]
\includegraphics[width=7cm]{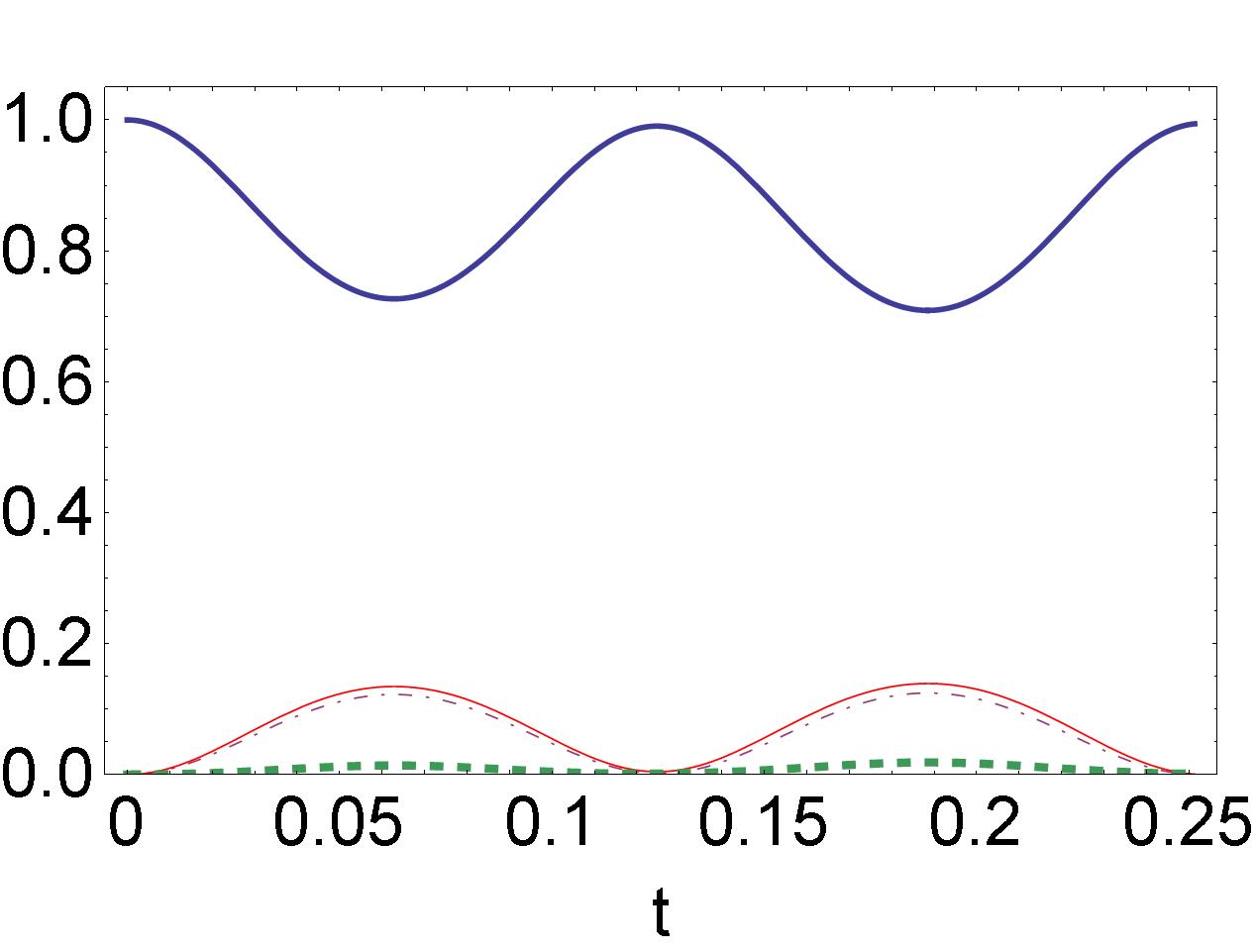} \caption{Same as Fig. \ref{figex1} in the \emph{nonadiabatic} case. The Hamiltonian
parameters are those given for Fig. \ref{figex1} except we set here
$q_{2}=7.33$. The initial state is again $\phi(x,0)=\psi_{n=2}^{\text{II}}(x,0)$.
The coefficients $\left\vert a_{k}(t)\right\vert ^{2}$ are shown
by the curves in thick solid blue ($k=2\text{), dot-dashed (\ensuremath{k=1\text{), solid red (\ensuremath{k=3\text{) and green dashed (\ensuremath{k=4}}}\ensuremath{)}}} }$.
The coefficients $a_{k}(t)$ for $k\protect\geq5$ are of smaller
and decreasing magnitude.}

\label{figex2} 
\end{figure}

For both examples, we take the initial state to be the ground state,
see Eq. (\ref{coeffci}) with $n=2$. The first illustration concerns
an instance for which the adiabatic approximation holds.\ This is
confirmed by the results obtained numerically for the coefficients
$a_{2}(t)$ and $a_{1,3}(t)$ plotted in Fig. \ref{figex1}: $\left\vert a_{2}(t)\right\vert ^{2}$
is almost always equal to unity for all times $0<t<T$, while $\left\vert a_{1}(t)\right\vert ^{2}$
is small and $\left\vert a_{3}(t)\right\vert ^{2}$ (as well as $\left\vert a_{k}(t)\right\vert ^{2}$ $\text{with }k>3$)
are negligible. In the second illustration, we keep the same parameters
as for the first illustration except for the amplitude of the oscillating
boundary, $q_{2}$, that is significantly increased. While $\left\vert a_{2}(t)\right\vert ^{2}$
is still the dominant term, $\left\vert a_{1}(t)\right\vert ^{2}$
and $\left\vert a_{3}(t)\right\vert ^{2}$ are not negligible, as
seen from the numerical results shown in Fig. \ref{figex2}.

\begin{figure}
\includegraphics[width=7cm]{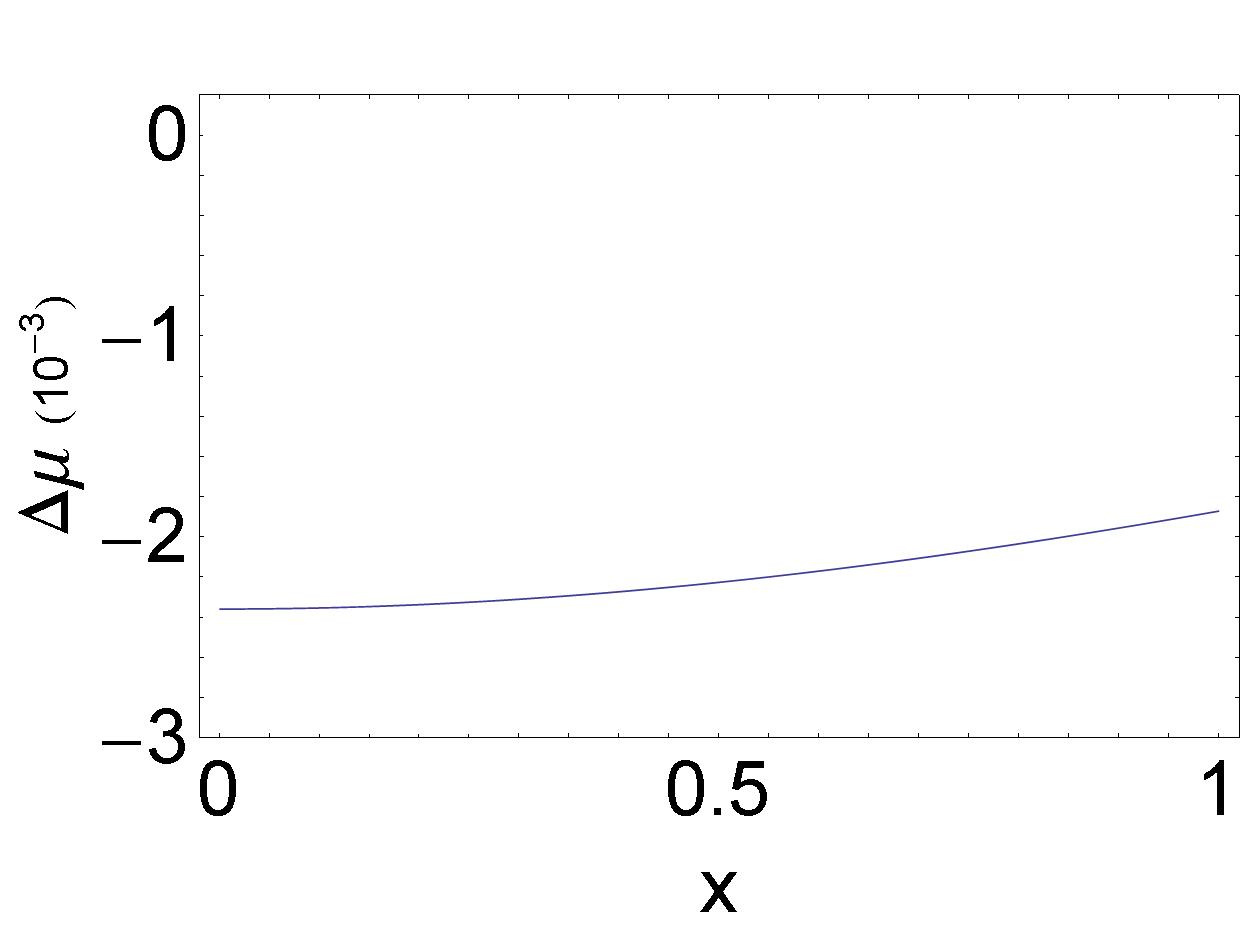}

\includegraphics[width=7cm]{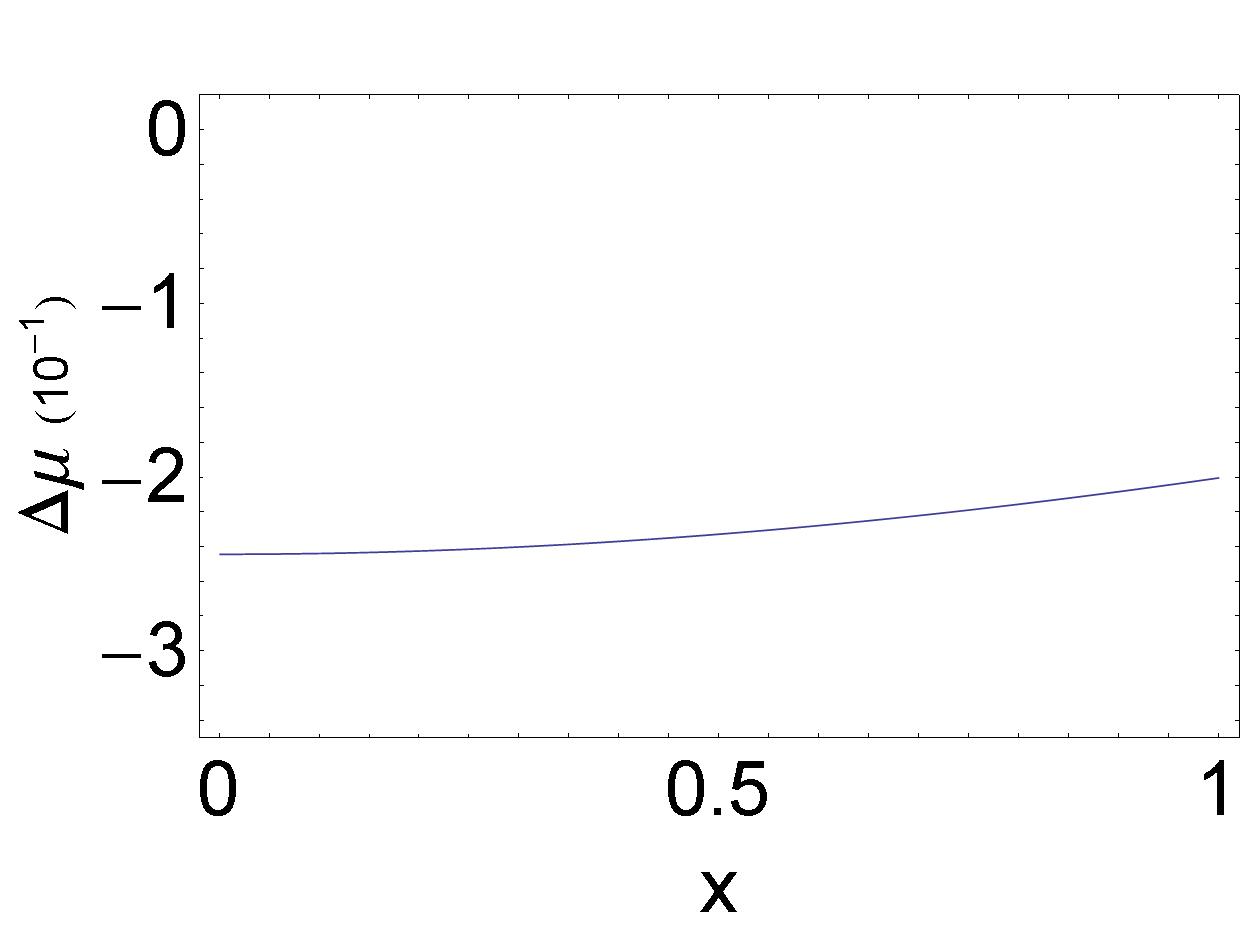}

\caption{Phase difference $\Delta\mu=\mu^{\text{III}}-\mu^{\text{I}}$ at $t=T$
between the case III and the case I cavities obtained from numerical
computations. Top: the case III cavity is in the adiabatic regime
(corresponding to the parameters given in Fig. \ref{figex1}). Applying
the analytical expression (\ref{muill}), the phase difference is
constant and computed to be $\mu^{ad}-\mu^{\text{I}}=-2.2\times10^{-3}.$
Bottom: the case III cavity is in a nonadiabatic regime (the parameters
are those given in Fig. \ref{figex2}). \label{figex3}}
\end{figure}

The phase difference $\Delta\mu=\mu^{III}-\mu^{I}$ at $t=T$ between
the case III and the case I cavities is shown in Fig. \ref{figex3}.
Recall that in case III, Eq. (\ref{fazeq}) is not exactly verified
-- it holds approximately when $q_{1}\approx q_{2}$ (and in particular
when the adiabatic approximation holds) but $\mu^{\text{III}}$ varies
with $x$ as $q_{1}$ becomes significantly different from $q_{2}$.
For our present purposes the important feature is that $\Delta\mu$
varies slowly with $x$ in the spatial region in which the phase difference
will be measured (i.e., in the vicinity of $x=0)$. This is the case
in the illustrations shown here.

A crucial feature for our argument on nonlocality concerns the absence
of superluminal velocities. As we have already mentioned, the Schr\"odinger
equation does not impose any bounds on energy eigenstates and admits
solutions of arbitrarily high energies, hence involving superluminal
velocities. We therefore need to check that no such states are needed
in order to account for the effect we observe in our illustrations.
From a numerical standpoint, it is straightforward to compute the
average velocity and its standard deviation as a function of time
and check it lies below the light velocity. As a rule of thumb, each
function $\psi_{n}(x,t)$ of Eq. (\ref{eig}) has a velocity component
obtained by decomposing the standing wave $\sin\frac{n\pi x}{L(t)}$
into the forward and backward traveling waves. The average velocity
of, say the forward traveling wave $\psi_{n}^{+}(x,t)$, is obtained
as, 
\begin{equation}
<v_{n}>=\frac{\left\langle \psi_{n}^{+}\right|P\left|\psi_{n}^{+}\right\rangle }{m}=\frac{n\pi\hbar}{mL(t)}+\frac{1}{2}\partial_{t}L(t).
\end{equation}
Alternatively, $<v>$ and $<v^{2}-<v>^{2}>^{1/2}$ can be obtained
straightforwardly for the standing wave $\psi_{n}(x,t)$. In both
cases, when time-averaged over a period this gives a velocity of the
order of $n\pi\hbar/m\left\langle L\right\rangle $, where $\left\langle L\right\rangle $
is the average cavity length over a period. We therefore need to ensure
only modes for which $n\pi\hbar/m\left\langle L\right\rangle \ll c$
contribute, a condition apparently fulfilled by large masses and long
cavities. But on the other hand in order to obtain a non-negligible
phase difference between cases I and III, we need $n$ to be large,
and $m$ and $\left\langle L\right\rangle $ to be small (see Sec.
\ref{subsec:B2}).

\begin{figure}[tb]
\includegraphics[width=7cm]{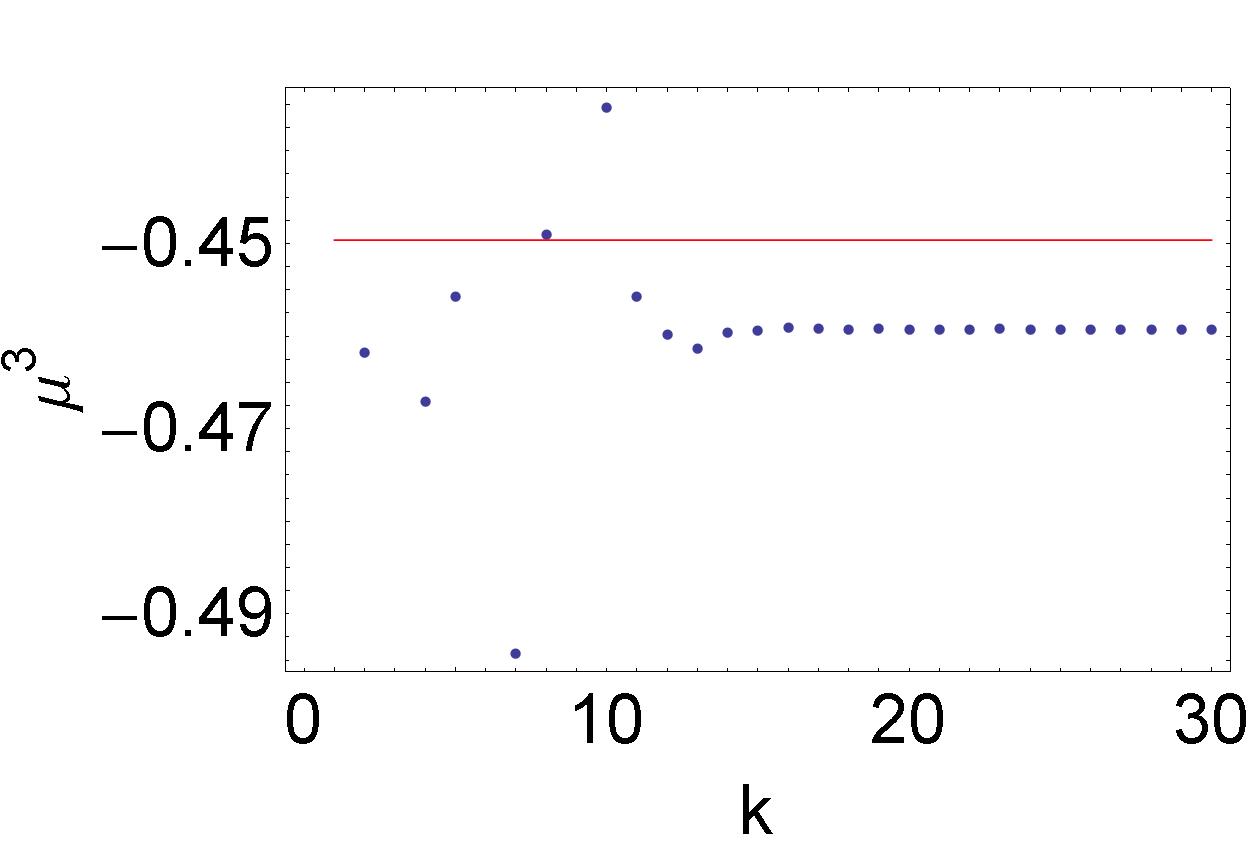}

\caption{The phase $\mu^{\text{III}}$ (blue dots) in the nonadiabatic case
of Fig. \ref{figex2} obtained from the numerically computed wavefunction
$\phi(x,t)$ as the basis size increases, cf. Eq. (\ref{exp}). It
is seen that states $\psi_{k}^{\text{II}}$ up to at least $k=15$
are necessary in order for the computed value of $\mu^{\text{III}}$
to converge. The solid red line represents $\mu^{\text{I}}$, the
phase increment in case I. The quantity that displays nonlocality
here is the phase difference between cases III and I (the blue dots
and the red line), $\Delta\mu=\mu^{\text{III}}-\mu^{\text{I}}$. }
\label{figfazbup} 
\end{figure}

We have seen that in the adiabatic case working with the sole function
$\psi_{n_{0}}^{\text{II}}(x,t)$ is sufficient to account for the
phase difference $\Delta\mu$. In the example given in Fig. \ref{figex1},
we have $n_{0}=2$ and $v/c\approx0.09$. In the nonadiabatic example,
more basis states need to be included; rather than setting a cut-off
value in the sum (\ref{exp}) as a function of the value of $\left|a_{k}(t)\right|$,
we use a stricter criterion requiring that the numerically computed
phase $\mu^{\text{III}}$ displays variations negligible compared
to $\Delta\mu$ and becomes constant as additional basis states are
included in the expansion. This is illustrated in Fig. \ref{figfazbup},
where it is seen that states up to $k=15$ must be included in the
expansion (\ref{exp}); this state corresponds to a velocity $v/c=15\pi\hbar/mcL_{0}\approx0.70$
(note however that for the overall state $\phi(x,t),$the time averages
over a period for $<v>$ and $<v^{2}-<v>^{2}>^{1/2}$ are much
lower, resp. $-0.001c$ and $0.064c$). Note that the non-adiabatic
regime is characterized by the fact that the non-diagonal terms in
Eq. (\ref{coefft}) contribute. As we showed in \ref{subsec:B2},
these non-diagonal terms are expected to increase for large values
of $L$ and when the time-dependent frequencies $\partial_{t}^{2}L_{1}(t)/L_{1}(t)$
and $\partial_{t}^{2}L_{2}(t)/L_{2}(t)$ become significantly different.
Therefore in order to avoid the contribution of high energy modes
(leading to superluminal velocities), we need rather large cavity
lengths, and hence we see that the non-adiabatic regime will be generic
and the adiabatic case an exception. In our numerical example however,
we have chosen a low value of $L$ and only changed the value of the
oscillating wall frequencies (through the values of the parameter
$q_{2}$) in order to illustrate the adiabatic and non-adiabatic cases
employing the same cavity except for the oscillation amplitude.

\section{Discussion}

We have investigated a system in which different boundary conditions
induce different cyclic global phases on the total wavefunction, although
the system evolves in identical potentials except near the boundary.
This global phase is acquired by the entire wavefunction, although part of the phase generation takes place at the boundary. In this sense it
appears to be a nonlocal effect. To observe a nonlocal signal, we perform an interference experiment wherein this effect produces a nonlocal relative phase instead of a global phase.

We assume this appears possible only because of some artifact of non-relativistic
quantum mechanics; it is indeed well-known that the Schr\"odinger equation
admits eigenstates of arbitrarily high energies, implying arbitrarily
high velocities. In a recent work \citep{mmw18} a similar type of
nonlocality in a related system (a cavity in free space expanding
linearly in time) was seen on the current density. Signaling could
be obtained by making weak measurements of the momentum. Discarding
the contribution of high energy basis states (that would propagate
superluminally) was somewhat subtle: several hundred basis states
needed to be included in order to compute the current density, and
while a basis restricted to subluminal states achieved the nonlocal
effect, convergence of the current density demanded the inclusion
of a more complete basis, containing superluminal states.

Here instead nonlocality arises through a periodic evolution of the
phase. Weak measurements are not required to observe signaling. Only
a very low number of basis states are involved in the computation,
and in the adiabatic limit, only a single state contributes. Nevertheless,
although it looks unlikely that high energy superluminal states are
reponsible for the form of nonlocality we have investigated here,
it remains impossible to totally discard their role. Indeed, from
a formal standpoint a complete basis in a non-relativistic setting
will necessarily include superluminally propagating states. And the
adiabatic limit is of course an approximation.

A second question that comes up concerns possible specific features
in the system employed. Indeed the present system combines two features,
each of which is known to lead to difficulties. First the confining
potential is modeled as an infinite wall, whereas inside the cavity
the potential is a time-dependent oscillator. Introducing infinite
discontinuities is known to lead to peculiar dynamical features \citep{aslangul}\citep{robbinett}.
It should be noted however that in the static cases (the standard
particle in a box problem or a harmonic oscillator confined by static
walls \cite{rau}) these peculiarities are not known to lead to any
unphysical results. Second, dealing with time-dependent boundary conditions
involves formally \citep{mosta1999,A2018} a different Hilbert space
at each time $t$. The time-dependent unitary mapping to a standard
problem -- a problem with fixed boundary conditions defined in a
single Hilbert space -- yields a Hamiltonian with a time-dependent
mass, so that the (local) boundary conditions are mapped to a (delocalized)
time-dependent parameter. We remark that mapping the time-dependence
of parameters in the original system to ther parameters in the transformed
system is a generic mathematical property of time-dependent unitary
transformations, and this mapping has in general no bearing on the
physical properties of the original system.

Conversely, one may wonder whether the nonlocal effect mentioned here
could be generic, in the sense that the wavefunction phase has a global
character even if it includes dynamical effects due to a potential
that is non-vanishing only in a small region over which the wavefunction
is defined. This would be analogous to the nonlocal features of the
Aharonov-Bohm phase \citep{AharonovAB,popescu-DNL,anandan88}. However
the AB effect is significantly different from the features characterizing
the present problem: it is an electromagnetic effect, the vector potential
is non-vanishing over the entire region in which the wavefunction
is defined, the individual phases are gauge-dependent, and there is
no signaling \citep{kampen,moulopoulos}. 

To conclude, we have investigated the evolution of the phase of the
wavefunction of a particle trapped in a confined time-dependent oscillator
with a moving boundary and found the phase to be nonlocal. We have
also seen that a protocol based on phase interference detection can
give rise to signaling, an apparent nonphysical artifact. Further
work is needed to pinpoint the different sources accounting for this
behavior within the formalism and to see if it is possible to discriminate
the effects due to dynamical nonlocality from the artifacts leading
to signaling.

\emph{Acknowledgments.} This research was supported (in part) by the
Fetzer Franklin Fund of the John E. Fetzer Memorial Trust.

\end{document}